\documentclass[twocolumn,prd,floatfix,showpacs,preprintnumbers,nofootinbib,%
superscriptaddress]{revtex4-1}

\usepackage{graphicx} 
\usepackage{amsmath,amssymb}
\usepackage{slashed}
\usepackage{xcolor}
\definecolor{lcolor}{rgb}{0.5,0,0}
\definecolor{citcolor}{rgb}{0,0.3,0.0}
\usepackage[breaklinks,colorlinks,urlcolor=blue,citecolor=citcolor,linkcolor=lc
olor]{hyperref}

\def\P{{\boldsymbol P}}
\def\k{{\boldsymbol k}}
\def\l{{\boldsymbol l}}
\def\p{{\boldsymbol p}}
\def\q{{\boldsymbol q}}

\newcommand{\der}{\mathrm{d}}
\newcommand{\xt}{{{\boldsymbol x}_\perp}}
\newcommand{\yt}{{{\boldsymbol y}_\perp}}
\newcommand{\bt}{{{\boldsymbol b}_\perp}}
\newcommand{\rt}{{{\boldsymbol r}_\perp}}
\newcommand{\kt}{{\k_\perp}}
\newcommand{\lt}{{\l_\perp}}
\newcommand{\pt}{{\p_\perp}}
\newcommand{\qt}{{\q_\perp}}
\newcommand{\Pt}{{\P_\perp}}

\newcommand{\ud}{\, \mathrm{d}}
\newcommand{\tr}{\, \mathrm{Tr} \, }
\newcommand{\nc}{{N_\mathrm{c}}}
\newcommand{\da}{d_\mathrm{A}}

\newcommand{\qso}{Q_\mathrm{s0}}

\newcommand{\lqcd}{\Lambda_{\mathrm{QCD}}}
\newcommand{\as}{\alpha_{\mathrm{s}}}

\newcommand{\Jpsi}{{J/\psi}}

\def\figscale{1.25}

\begin{document}

\title{Forward $J/\psi$ production at high energy: centrality dependence and 
mean transverse momentum}
\author{B. Duclou\'e}
\affiliation{
Department of Physics, University of Jyv\"askyl\"a %
 P.O. Box 35, 40014 University of Jyv\"askyl\"a, Finland
}
\affiliation{
Helsinki Institute of Physics, P.O. Box 64, 00014 University of Helsinki,
Finland
}

\author{T. Lappi}
\affiliation{
Department of Physics, University of Jyv\"askyl\"a %
 P.O. Box 35, 40014 University of Jyv\"askyl\"a, Finland
}
\affiliation{
Helsinki Institute of Physics, P.O. Box 64, 00014 University of Helsinki,
Finland
}

\author{H. M\"antysaari}
\affiliation{Physics Department, Brookhaven National Laboratory, Upton, NY 11973, USA}

\pacs{
13.85.Ni,   
14.40.Pq,  
24.85.+p,       
25.75.Cj  
}

\begin{abstract}
Forward rapidity $\Jpsi$ meson production  in proton-nucleus 
collisions can be an important constraint of descriptions of the small-$x$ 
nuclear wavefunction. 
In an earlier work we studied this process using a dipole cross section
satisfying the Balitsky-Kovchegov equation, fit to HERA inclusive data and
consistently extrapolated to the nuclear case using a standard Woods-Saxon distribution.
In this paper we present further  calculations of these cross sections,
studying the mean transverse momentum of the meson and the dependence on collision centrality. We also extend the calculation to backward rapidities using nuclear parton distribution functions.
We show that the parametrization is overall rather consistent with the available 
experimental data. However, there is a tendency towards a too strong centrality
dependence. This can be traced back to the rather small transverse area occupied by 
small-$x$ gluons in the nucleon that is seen in the HERA data, compared to the total
inelastic nucleon-nucleon cross section.
\end{abstract}

 \maketitle

\section{Introduction}

The production of $J/\psi$ mesons at forward rapidity in high energy proton-proton and proton-nucleus collisions can provide valuable information on gluon saturation. Indeed, the production of particles at forward rapidity probes the target at very small $x$, where saturation effects should be enhanced. In particular, the charm quark mass being of the same order of magnitude as the saturation scale, $J/\psi$ production should be sensitive to these dynamics. The charm quark mass is also large enough to provide a hard scale, making a perturbative study of this process possible. In addition, $J/\psi$ production has been the subject of many experimental studies at the LHC, both in proton-proton~\cite{Khachatryan:2010yr,Aaij:2011jh,Aad:2011sp,Chatrchyan:2011kc,Abelev:2014qha,Aaij:2015rla,Adam:2015rta} and in proton-nucleus~\cite{Abelev:2013yxa,Aaij:2013zxa,Adam:2015iga,Aad:2015ddl,Adam:2015jsa} collisions. This provides a lot of data to confront with nuclear effects predicted by various theoretical models,
both in the Color Glass Condensate (CGC) framework \cite{Fujii:2013gxa,Ducloue:2015gfa,Ma:2015sia,Watanabe:2015yca,Fujii:2015lld} and
as constraints for nuclear parton distribution functions \cite{Albacete:2013ei}
and energy loss in cold nuclear matter~\cite{Vogt:2010aa,Arleo:2012rs,Arleo:2014oha,Vogt:2015uba}.

In a recent work~\cite{Ducloue:2015gfa} we studied, in the CGC framework, the production of forward $J/\psi$ mesons in proton-proton and proton-nucleus collisions at the LHC. We showed that, when using the Glauber approach to generalize the dipole cross section to nuclei, the nuclear suppression for minimum bias events is smaller than in previous CGC calculations such as~\cite{Fujii:2013gxa}, and much closer to experimental data. In this paper we will study, in the same framework, other observables of interest in this process, such as $J/\psi$ production at backward rapidity, the centrality dependence in the optical Glauber model 
and the mean transverse momentum of the produced $\Jpsi$'s.

\section{Formalism}

Let us briefly recall the main steps of the calculation. For more details we refer the reader to Ref.~\cite{Ducloue:2015gfa}. We use the color evaporation model (CEM) to relate the cross section for $J/\psi$ production to the $c\bar{c}$ pair production cross section. In this model a fixed fraction $F_{J/\psi}$ of the $c\bar{c}$ pairs produced below the $D$-meson threshold are assumed to hadronize into $J/\psi$ mesons:
\begin{equation} 
\frac{\ud\sigma_{J/\psi}}{\ud^2\Pt\ud Y}
=
F_{J/\psi} \; \int_{4m_c^2}^{4M_D^2} \ud M^2
\frac{\ud\sigma_{c\bar c}}
{\ud^2\Pt \ud Y \ud M^2}
\, ,
\label{eq:dsigmajpsi}
\end{equation}
where $\Pt$, $Y$ and $M$ are the transverse momentum, the rapidity and the invariant mass of the $c\bar{c}$ pair respectively, $M_D=1.864$ GeV is the D meson mass and $m_c$ is the charm quark mass that we will vary between 1.2 and 1.5 GeV. Note that in this work we will focus on ratios where $F_{J/\psi}$ cancels so we do not need to fix it to any specific value here.

The study of gluon and quark pair production in the dilute-dense limit of the CGC formalism was started some time ago~\cite{Blaizot:2004wu,Blaizot:2004wv} (see also \cite{Kharzeev:2012py}) and used in several calculations such as~\cite{Fujii:2006ab,Fujii:2005rm,Fujii:2013gxa,Fujii:2013yja}. The physical picture is the following: an incoming gluon from the projectile can split into a quark-antiquark pair either before or after the interaction with the target. The partons propagating trough the target are assumed to interact eikonally with it, picking up a Wilson line factor in either the adjoint (for gluons) or the fundamental (for quarks) representation. Since we study $J/\psi$ production at forward rapidity, where the projectile is probed at large $x$, we will use the collinear approximation in which the incoming gluon is assumed to have zero transverse momentum. In this approach the cross section for $c\bar c$ production reads, in the large-$\nc$ limit~\cite{Fujii:2013gxa},
\begin{multline}
\frac{\ud \sigma_{c\bar{c}}}{\ud^2\pt \ud^2\qt \ud y_p \ud y_q}
= 
\frac{\as^2 \nc}{8\pi^2 \da }
\\ \times
\frac{1}{(2\pi)^2}
\int\limits_{\kt}
\frac{\Xi_{\rm coll}(\pt + \qt,\kt)}{(\pt + \qt)^2}
x_1 G_p(x_1,Q^2) 
\\
\times
\phi_{y_2=\ln{\frac{1}{x_2}}}^{q\bar{q},g}(\pt + \qt,\kt)
\;
 ,
\label{eq:dsigmaccbarcoll}
\end{multline}
where $\pt$ and $\qt$ denote the transverse momenta of the quarks, $y_p$ and $y_q$ their rapidities, $\int_{\kt} \equiv \int \ud^2 \kt / (2\pi)^2$,
and $\da\equiv \nc^2-1$ is the dimension of the adjoint representation of
SU($\nc$). The longitudinal momentum fractions $x_1$ and $x_2$ probed in the projectile and the target respectively are given by
\begin{equation}
x_{1,2}=\frac{\sqrt{\Pt^2+M^2}}{\sqrt{s}}e^{\pm Y} \; .
\end{equation}
The explicit expression for the ``hard matrix element'' $\Xi_{\rm coll}$ is given in Ref.~\cite{Fujii:2013gxa}.
In Eq.~(\ref{eq:dsigmaccbarcoll}), $G_p(x_1,Q^2)$ is the gluon density in the probe and is described, in the collinear approximation that we use here, in terms of usual parton distribution functions (PDFs). In the following we use, unless otherwise stated, the MSTW 2008~\cite{Martin:2009iq} parametrization for  $G_p$. For consistency we use the leading other (LO) parton distribution  since also the rest of the calculation is made at this order. To estimate the uncertainty associated with the choice of the factorization scale $Q$, we will vary it between $\frac{1}{2}\sqrt{\Pt^2+M^2}$ and $2\sqrt{\Pt^2+M^2}$.

The function $\phi_{_Y}^{q \bar{q},g}$ describes the propagation of a $q\bar{q}$ pair in the color field of the target and reads
\begin{multline}\label{eq:defphi}
\phi_{_Y}^{q \bar{q},g}(\lt,\kt)=
\int\der^2 \bt \frac{N_c\lt^2}{4 \as} \; 
\\ \times
S_{_Y}(\kt,\bt) \;
S_{_Y}(\lt-\kt,\bt) \;,  
\end{multline}
where $\bt$ is the impact parameter.
In this expression, the function $S_{_Y}(\kt,\bt)$ contains all the information about the target. It is the fundamental representation dipole correlator:
\begin{equation}
S_{_Y}(\kt,\bt) = \int \ud^2 \rt  
e^{i\kt \cdot \rt }
S_{_Y}(\rt,\bt) \;,
\end{equation}
with
\begin{equation}
S_{_Y}(\xt-\yt,\bt) = \frac{1}{\nc }\left< \tr U^\dag(\xt)U(\yt)\right>,
\end{equation}
where $U(\xt)$ is a fundamental representation Wilson line in the color field of the target.
The dipole correlator $S_{_Y}(\kt,\bt)$ is obtained by solving numerically the running coupling Balitsky-Kovchegov (rcBK) equation~\cite{Balitsky:1995ub,Kovchegov:1999ua,Balitsky:2006wa}. For the initial condition we use the 'MV$^e$' parametrization introduced in Ref.~\cite{Lappi:2013zma}, which reads, in the case of a proton target,
\begin{multline}\label{eq:icp}
S^p_{Y= \ln \frac{1}{x_0}}(\rt,\bt) = \exp \bigg[ -\frac{\rt^2 \qso^2}{4} 
\\ \times
\ln \left(\frac{1}{|\rt| \lqcd}\!+\!e_c \cdot e\right)\bigg],
\end{multline}
where $x_0=0.01$. The running coupling is taken as:
\begin{equation}
 \as(r) = \frac{12\pi}{(33 - 2N_f) \log \left(\frac{4C^2}{r^2\lqcd^2} \right)} \;,
\end{equation}
where $C$ parametrizes the uncertainty related to the scale of the strong coupling in the transverse coordinate space.

The free parameters in these expressions are obtained by fitting the combined inclusive HERA DIS cross section data~\cite{Aaron:2009aa} for $Q^2<50$ GeV$^2$ and $x<0.01$. Their best fit values (with $\chi^2/\text{d.o.f} = 1.15$) are $\qso^2= 0.060$ GeV$^2$, $C^2= 7.2$, $e_c=18.9$  and $\sigma_0/2 = 16.36$ mb.
In the case of a proton target, the dipole amplitude does not have an explicit impact parameter dependence and we thus make the replacement
\begin{equation}\label{eq:defsigma0}
\int\der^2 \bt \to \frac{\sigma_0}{2}
\end{equation}
in Eq.~(\ref{eq:defphi}), $\frac{\sigma_0}{2}$ corresponding to the effective proton transverse area measured in DIS experiments.
The fit only includes light quarks. In particular this leaves the charm quark mass 
that would be consistent with the DIS data in this model still uncertain, which is why we vary it in a rather large range for the uncertainty estimate in this work.

\begin{figure*}[tbp]
	\centering
	\includegraphics[scale=\figscale]{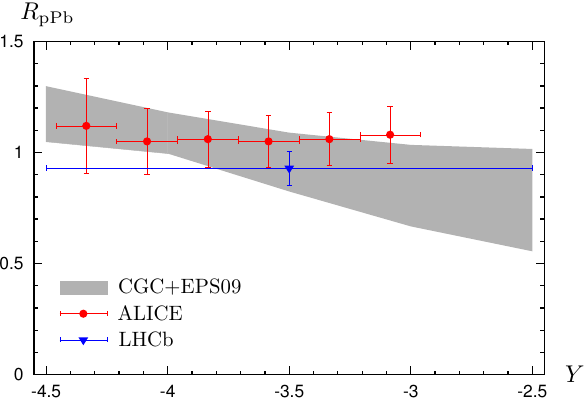}
	\hspace{0.5cm}
	\includegraphics[scale=\figscale]{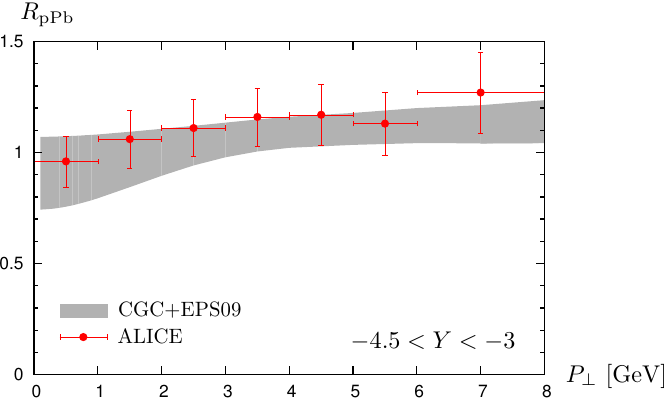}
	\caption{Nuclear modification factor $R_\text{pPb}$ at negative rapidity as a function of $Y$ (left) and $P_\perp$ (right) at $\sqrt{s_{NN}}=5$ TeV. Data from Refs.~\cite{Abelev:2013yxa,Aaij:2013zxa,Adam:2015iga}.}
	\label{fig:RpA_backward}
\end{figure*}

\begin{figure*}[tbp]
	\centering
	\includegraphics[scale=\figscale]{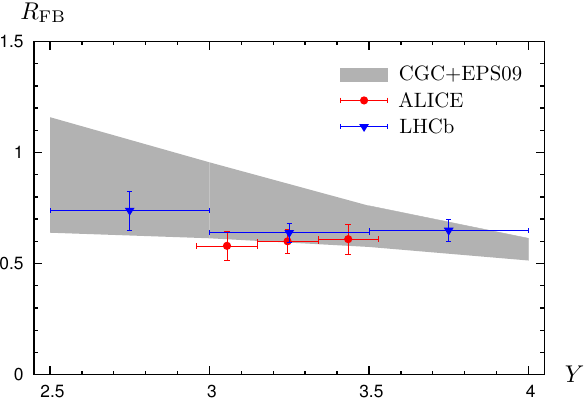}
	\hspace{0.5cm}
	\includegraphics[scale=\figscale]{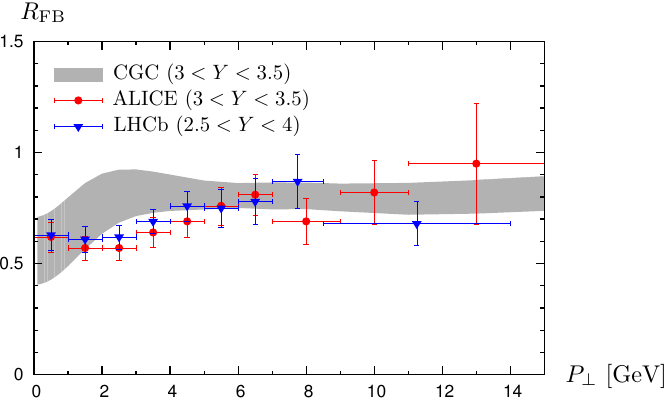}
	\caption{Forward to backward ratio in proton-lead collisions as a function of $Y$ (left) and $P_\perp$ (right) at $\sqrt{s_{NN}}=5$ TeV. Data from Refs.~\cite{Abelev:2013yxa,Aaij:2013zxa}.}
	\label{fig:RFB}
\end{figure*}

To generalize this proton dipole correlator to the case of a nuclear target  we use, as in~\cite{Lappi:2013zma}, the optical Glauber model. In this model the
gluons at the initial rapidity $Y=\ln 1/x_0$ are localized in the individual nucleons
of the nucleus. The nucleons are then taken to be distributed randomly and independently in the transverse plane according to the standard Woods-Saxon nuclear density profile. An analytical average over the positions of the nucleons leads to the following  initial condition for the rcBK evolution of a nuclear target:
\begin{multline}\label{eq:ica}
S^A_{Y=\ln \frac{1}{x_0}}(\rt,\bt) = \exp\Bigg[ -A T_A(\bt) 
\frac{\sigma_0}{2} \frac{\rt^2 \qso^2}{4} 
\\
\times
\ln \left(\frac{1}{|\rt|\lqcd}+e_c \cdot e\right) \Bigg] \; .
\end{multline}
Here $T_A$ is the standard Woods-Saxon transverse thickness function of the nucleus:
\begin{equation}
T_A(\bt)= \int dz \frac{n}{1+\exp \left[ \frac{\sqrt{\bt^2 + z^2}-R_A}{d} \right]} \; ,
\end{equation}
with $d=0.54\,\mathrm{fm}$ and $R_A=(1.12A^{1/3}-0.86A^{-1/3})\,\mathrm{fm}$, and $n$ is defined such that $T_A$ is normalized to unity. All the other parameters in the initial condition (\ref{eq:ica}), which is evolved using the rcBK equation for each $\bt$, are the same as in the proton case.

In this model the dipole amplitude depends on the impact parameter and we need to integrate explicitly over it. The impact parameter dependence, which carries over to the centrality dependence, thus appears naturally in this model. Our practical procedure for carrying out this comparison will be discussed in more detail in 
Sec.~\ref{sec:centrality}.

\section{Backward rapidity}

In our previous work~\cite{Ducloue:2015gfa} we only considered $J/\psi$ production in proton-proton and proton-nucleus collisions at forward rapidity. In these kinematics the process can be seen as the collision of a dilute proton probed at large $x$, which can be described using well known parton distribution functions (PDFs), and a dense target described in terms of classical color fields. The nuclear modification of $J/\psi$ production was also measured at backward rapidity by ALICE~\cite{Abelev:2013yxa,Adam:2015iga} and LHCb~\cite{Aaij:2013zxa}. In this case the produced $J/\psi$ is moving in the direction of the incoming nucleus and the physical picture is the same as at forward rapidity, but with the roles of the projectile and the target interchanged, i.e. a dilute proton or nucleus interacts with a dense proton target. The latter is described in the same way as in proton-proton collisions at forward rapidity, while the projectile is described either by a PDF in the case of a proton or by a nuclear PDF (nPDF) in the case of a nucleus. Therefore the calculation is very similar to the case of proton-proton collisions in our previous study. Note, however, that in general nuclear PDFs are less tightly constrained by experimental data than usual proton PDFs. In the following we will use the leading order EPS09 nPDF parametrization~\cite{Eskola:2009uj} which provides additional error sets to estimate this uncertainty. The LO EPS09 analysis uses the CTEQ6L1~\cite{Pumplin:2002vw} proton PDFs as a reference, therefore for consistency we use same parametrization of proton PDFs when computing the nuclear modification factor at backward rapidity. Nevertheless, while it could be sizeable for the cross section, the difference compared to the  MSTW 2008 parametrization is very small for the nuclear
modification factor.

In Fig.~\ref{fig:RpA_backward} we show the nuclear modification factor $R_\text{pA}$, defined as
\begin{equation}
R_{\rm pA}= \frac{1}{A}\frac{\frac{\ud \sigma^\text{pA}}{\ud^2 \Pt \ud Y}}
{\frac{\ud\sigma^\text{pp}}{\ud^2 \Pt \ud Y}} \; ,
\label{eq:defrpa}
\end{equation} 
as a function of $Y$ and $P_\perp$ obtained in this way at negative rapidity compared with data from ALICE~\cite{Abelev:2013yxa,Adam:2015iga} and LHCb~\cite{Aaij:2013zxa} experiments. The uncertainty in our calculation is significantly larger than at forward rapidity~\cite{Ducloue:2015gfa}. This is due to the fact that we include in our uncertainty band, in addition to the variation of the charm quark mass and of the factorization scale, the nuclear PDF uncertainty obtained following the procedure described in Ref.~\cite{Eskola:2009uj}. In particular, our lower bound for the factorization scale is $Q=\frac{1}{2}\sqrt{\Pt^2+M^2}$ which can reach values of less than 2 GeV at small transverse momentum. The nuclear PDFs are not well constrained at such small scales at the moment. Nevertheless the general agreement with data is quite good taking into account the rather large theoretical and experimental uncertainties. In this case deviations of $R_\text{pPb}$ from unity are entirely due to the nuclear PDFs.

Now that we have computed the nuclear modification factor both at forward~\cite{Ducloue:2015gfa} and backward rapidities, we have access to the forward to backward ratio $R_\text{FB}$, defined as
\begin{equation}
R_\text{FB}(P_\perp,Y)=\frac{R_\text{pA}(P_\perp,Y)}{R_\text{pA}(P_\perp,-Y)} \; .
\end{equation}
This ratio can be interesting to study because there may be an additional cancellation of some uncertainties common to the numerator and the denominator. In particular, when determining the nuclear modification factor, experimental studies such as~\cite{Abelev:2013yxa,Aaij:2013zxa} have to use an interpolation for the reference proton-proton cross section since there is no data at $\sqrt{s}=5$ TeV. This interpolation is not needed to study $R_\text{FB}$, but the final statistical uncertainty may be larger if the coverage in rapidity by the detector is not symmetric with respect to 0. Concerning our calculation, we have seen that at negative rapidity the uncertainty on $R_\text{pA}$ due to nuclear PDFs is rather large. This error will remain in $R_\text{FB}$ since the computation of $R_\text{pA}$ at positive rapidity does not involve nPDFs. Indeed, we see from Fig.~\ref{fig:RFB}, where we show the forward to backward ratio as a function of $Y$ and $P_\perp$, that the uncertainty on this quantity is still quite large. Nevertheless, within this error band the agreement with data is reasonable, although the variation at low $P_\perp$ seems to be steeper than in the data. In Fig.~\ref{fig:RFB} (R) we only show our results for $R_\text{FB}$ as a function of $P_\perp$ integrated over the same $Y$ range as ALICE data~\cite{Abelev:2013yxa}, which is slightly smaller than for LHCb data~\cite{Aaij:2013zxa}, but our results for $2.5<Y<4$ would be very similar and ALICE and LHCb data are compatible with each other.

\begin{figure*}[tbp]
	\def\sca{1.15}
	\includegraphics[scale=\sca]{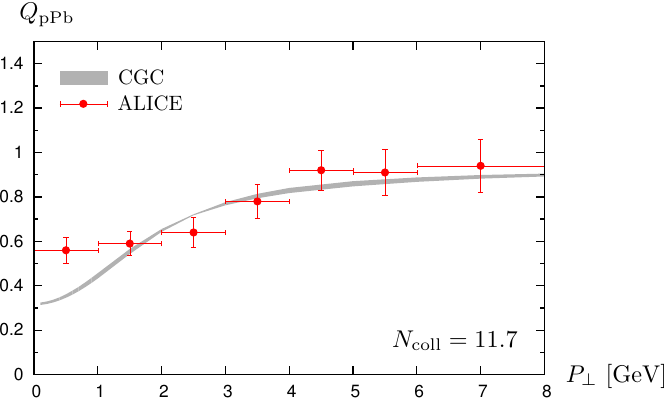}
	\hspace{0.5cm}
	\includegraphics[scale=\sca]{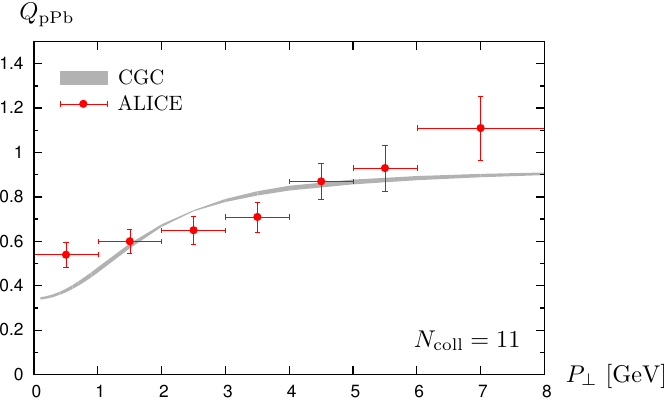}
	
	\vspace{0.3cm}
	\includegraphics[scale=\sca]{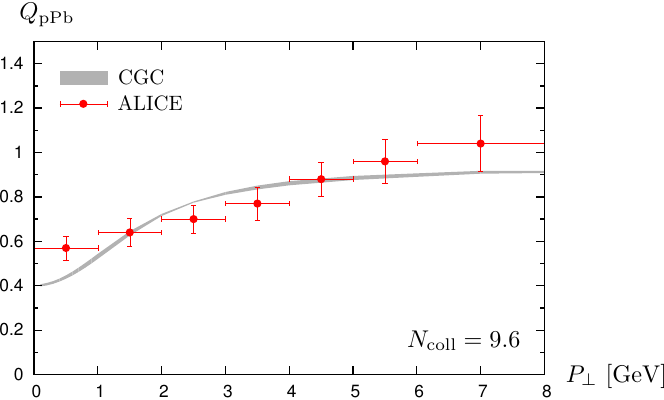}
	\hspace{0.5cm}
	\includegraphics[scale=\sca]{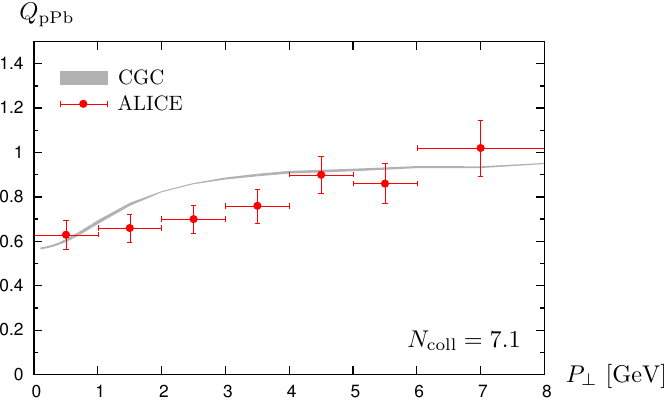}

	\vspace{0.3cm}
	\includegraphics[scale=\sca]{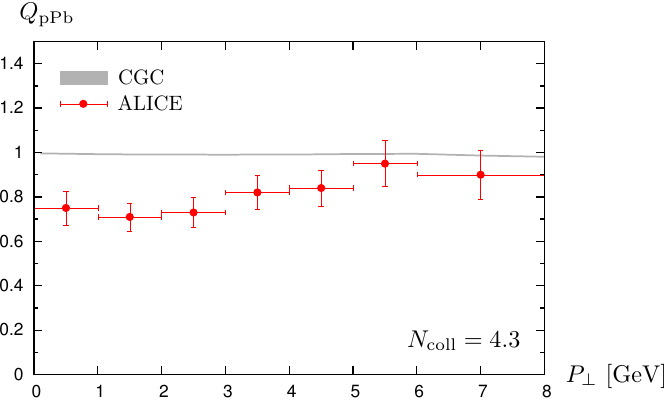}
	\caption{Nuclear modification factor $Q_\text{pPb}$ as a function of $P_\perp$ at $\sqrt{s_{NN}}=5$ TeV in different centrality bins compared with ALICE data~\cite{Adam:2015jsa}.}
	\label{fig:QpA_bins}
\end{figure*}

\begin{table}[tb]
	\centering
	\begin{tabular}{|c|c|c|c|}
		\hline
		Centrality class & $\langle N_\text{coll} \rangle_\text{opt.}$ & $\langle N_\text{coll} \rangle_\text{ALICE}$
		&$b$ [fm]
 \\
		\hline
		2--10\%   & 14.7 & $11.7 \pm 1.2 \pm 0.9$ &  4.14\\
		10--20\%  & 13.6 & $11.0 \pm 0.4 \pm 0.9$ & 4.44 \\
		20--40\%  & 11.4 & $9.6 \pm 0.2 \pm 0.8$ & 4.94 \\
		40--60\%  & 7.7  & $7.1 \pm 0.3 \pm 0.6$ & 5.64 \\
		60--80\%  & 3.7  & $4.3 \pm 0.3 \pm 0.3$ & 6.29  \\
		80--100\% & 1.5  & $2.1 \pm 0.1 \pm 0.2$ & 6.91\\
		\hline
	\end{tabular}
	\caption{Average number of binary collisions in each centrality class as obtained in the optical Glauber model compared with the value estimated by ALICE~\cite{Adam:2015jsa}. The values of $b$ in the last column are solved from the relation $N_\text{coll,opt.}(b)=\langle N_\text{coll} \rangle_\text{ALICE}$
}
	\label{tab:Ncoll}
\end{table}

\section{Centrality dependence}
\label{sec:centrality}

We have seen that the optical Glauber model contains an explicit impact parameter dependence which can be related to centrality determinations at experiments. In this section we will compare this centrality dependence with its recent measurement at the LHC by the ALICE collaboration in the range $2<Y<3.5$~\cite{Adam:2015jsa}.

\subsection{Centrality in the optical Glauber model}

We still need to relate the explicit impact parameter dependence in our model to the definition of centrality used by experiments. The experimental data are usually presented in terms of centrality classes. In the optical Glauber model these classes would be defined by calculating the impact parameter range corresponding to the centrality class $(c_1-c_2)\%$ using the relation
\begin{equation}
(c_1-c_2)\% = \frac{1}{\sigma_\text{inel}^{\text{pA}}} \int_{b_1}^{b_2} \der^2 \bt p(\bt).
\label{eq:centrality}
\end{equation}
Here $\sigma_\text{inel}^{\text{pA}}$ is the total inelastic proton-nucleus cross section, given by
\begin{equation}
\sigma_\text{inel}^{\text{pA}} = \int \der^2 \bt \, p(\bt) \; ,
\end{equation}	
and the scattering probability at impact parameter $\bt$ is
\begin{equation}
p(\bt) \approx 1- e^{-A T_A(\bt) \sigma_\text{inel}},
\end{equation}
where $\sigma_\text{inel}$ is the total inelastic nucleon-nucleon cross section. The particle yield in each centrality class is then given by
\begin{equation}
\frac{\der N}{\der^2 \Pt \der Y} = \frac{ \int_{b_1}^{b_2} \der^2 \bt \frac{\der N(\bt)}{\der^2 \Pt \der Y} } {\int_{b_1}^{b_2} \der^2 \bt \, p(\bt)},
\end{equation}
where the values of $b_1$ and $b_2$ are obtained from Eq.~(\ref{eq:centrality}) and $\frac{\der N(\bt)}{\der^2 \Pt \der Y}$ corresponds to the expression of the cross section before integrating over $\bt$.

In practice, however, this straightforward procedure cannot be used for comparing our calculation with the experimental centrality classes. This can be seen most clearly by  calculating the average number of binary nucleon-nucleon collisions for each centrality class.
In the optical Glauber model this is computed using the relation
\begin{equation}
\langle N_\text{coll} \rangle_\text{opt.} = \frac{\int_{b_1}^{b_2} \der^2 \bt N_\text{bin}(\bt)}{\int_{b_1}^{b_2} \der^2 \bt p(\bt)},
\end{equation}
with
\begin{equation}
	 N_\text{bin}(\bt) = A T_A(\bt) \sigma_\text{inel} \,.
\end{equation}
Table~\ref{tab:Ncoll} shows $\left< N_\text{coll} \right>$ for the optical Glauber centrality classes compared to the values given by ALICE~\cite{Adam:2015jsa} for the experimental ones.
For central collisions the average number of binary collisions estimated by ALICE is smaller than in the optical Glauber model, while the opposite is true for peripheral collisions. 

The impact parameter is not directly observable, so the experimental centrality selection has to use some other observable as a proxy for it. In the case of the ALICE analysis~\cite{Adam:2015jsa} the observable used is the energy of the lead-going side Zero-Degree Calorimeter  (ZDC). Due to the large fluctuations in the signal for a fixed impact parameter, the values of $N_\text{coll}$ vary less strongly with centrality in the experimental classes than in the optical Glauber ones. Thus, for example, a relatively peripheral smaller-$N_\text{coll}$ event can end up in the most central class in the case of a fluctuation in the ZDC signal.

\subsection{Fixed impact parameter approximation}

Developing a detailed Monte Carlo Glauber model that would enable us to exactly match the experimental centrality classes would be beyond the scope of this work. Therefore  we start from the assumption that the ALICE Glauber model is correct and produces a reliable estimate for the $\langle N_\text{coll} \rangle$ in each experimental class. We then assume that the mapping between this $\langle N_\text{coll} \rangle$ and the impact parameter is accurately enough described by our optical Glauber model, and use the relation $N_\text{coll. opt}(b)=\langle N_\text{coll} \rangle_\text{ALICE}$ to determine a mean impact parameter corresponding to the experimental centrality class. The values of $b$ resulting from this procedure are shown in Table~\ref{tab:Ncoll}.
We then calculate the nuclear modification factor for the centrality class using this fixed value of $b$.
We note that following this procedure for the 80-100\% centrality class would lead to a value of $b$ for which our calculation would not be applicable because the saturation scale of the nucleus falls below the one of the proton. Therefore in the following we will only show results for the five most central classes considered by ALICE. 

\begin{figure*}[tbp]
	\centering
	\includegraphics[scale=\figscale]{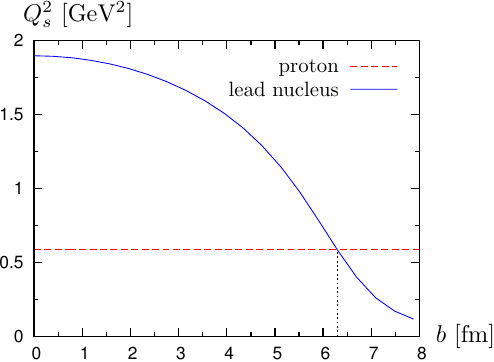}
	\hspace{2cm}
	\includegraphics[scale=\figscale]{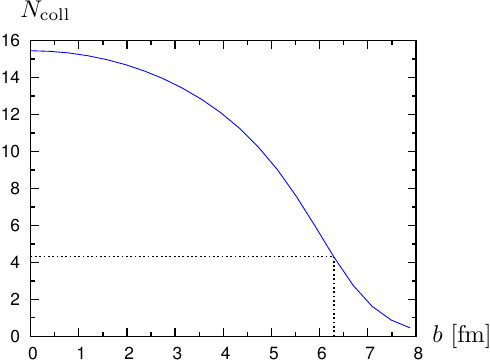}
	\caption{Left: saturation scale of the lead nucleus in the MV$^e$ parametrization as a function of the impact parameter compared with the saturation scale of the proton at $x=10^{-4}$. Right: number of binary collisions in proton-lead collisions as a function of the impact parameter in the optical Glauber model.}
	\label{fig:Qs2_Ncoll}
\end{figure*}

In Fig.~\ref{fig:QpA_bins} we show a comparison of our calculation with ALICE data for the nuclear modification factor in different centrality classes, $Q_\text{pA}$, defined as
\begin{equation}
Q_{\rm pA}= \frac{\frac{\ud N^\text{pA}}{\ud^2 \Pt \ud Y}}
{ A \langle T_A \rangle \frac{\ud\sigma^\text{pp}}{\ud^2 \Pt \ud Y}} \; ,
\end{equation}
as a function of $P_\perp$. We observe that the description of experimental data is generally satisfactory in the first four bins. For the fifth bin the value of $Q_\text{pPb}$ obtained in the optical Glauber model is almost constant and very close to one, while the data still shows a significant variation with $P_\perp$.

The discrepancy with experimental data for peripheral collisions comes from the fact that in our model the saturation scale of the lead nucleus falls below the one of the proton for a value of $b$ of the order of 6.3 fm, as shown in Fig.~\ref{fig:Qs2_Ncoll}~(L). We see from Fig.~\ref{fig:Qs2_Ncoll}~(R), where we show the number of binary collisions as a function of $b$, that this corresponds to $N_\text{coll} \sim 4.3$. This is the point where, by definition, $Q_\text{pA}$ reaches 1 in our calculation and beyond which the validity of the framework we have used is questionable. On the other hand, ALICE data shows that $Q_\text{pA}$ is still significantly smaller than 1 down to $N_\text{coll} \sim 2.1$ (see Fig.~\ref{fig:QpA_Ncoll}).

The strong centrality dependence is caused by the value of $\frac{\sigma_0}{2}$ extracted from DIS fits being much smaller than the total inelastic nucleon-nucleon cross section. 
The HERA data, both the inclusive cross section fitted  in Ref.~\cite{Lappi:2013zma} and data on exclusive vector meson production
(see e.g.~\cite{Chekanov:2004mw}), lead to a picture where the small-$x$ gluons that can participate in a hard process in a proton are concentrated in a rather small area in the transverse plane (see also the discussion in Ref.~\cite{Frankfurt:2010ea}). This small-$x$ gluon ``hot spot'' is then surrounded by a larger ``cloud'' that only participates in soft interactions, contibuting to the total nucleon-nucleon inelastic cross section.
Our model takes this picture to the extreme, by assuming that at the initial rapidity $\ln 1/x_0$ the gluons contributing to $\Jpsi$ production are concetrated in the area 
$\sigma_0/2 \sim 0.3 \sigma_\text{inel}$ inside the target nucleons. Thus, for peripheral collisions, the probe proton can overlap with the soft cloud of $N_\text{coll} \sim 4$ target nucleons while still seeing on average only one small-$x$ gluon hot spot in the target, thus 
leaving a hard process like $\Jpsi$ production approximately unmodified.

This centrality dependence could probably be mildened by using a larger value for $\frac{\sigma_0}{2}$ of the order of $\sigma_\text{inel}$ (as is effectively done in \cite{Fujii:2013gxa}). This would, however, lose the consistency of our description of the nucleon from HERA to the LHC. Also, as discussed in Refs.~\cite{Lappi:2013zma,Ducloue:2015gfa}, varying these parameters in an uncontrolled way could very easily, depending on how exactly it is done, lead to an excessive suppression for minimum bias collisions, or to an $R_\text{pA}$ that is very far from unity even at high transverse momentum.

Note that the saturation scale (or gluon density) at the edge of the nucleus being smaller than that of the proton is an artefact of averaging over the transverse locations of the dense but small gluon hot spots of the nucleons in the target nucleus. Therefore we do not use the optical Glauber parametrization in this region, but explicitly set $R_\text{pA}$ to unity. Nevertheless, even an explicit Monte Carlo Glauber procedure with the same parameters  would not change the ordering $\sigma_0/2 < \sigma_\text{inel}$ that leads to the absence of nuclear effects for peripheral collisions with  $N_\text{coll} \lesssim 4$. 

\begin{figure}[tb]
	\centering
	\includegraphics[scale=\figscale]{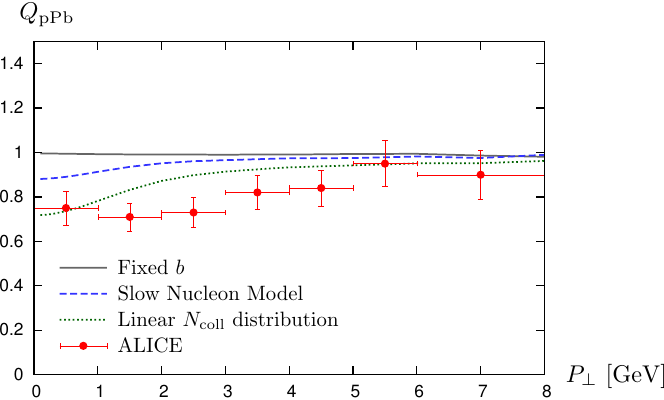}
	\caption{$Q_\text{pPb}$ at a function of $P_\perp$ at $\sqrt{s_{NN}}=5$ TeV in the 60-80\% centrality class, both when using a fixed impact parameter and when integrating explicitly over $b$ using the $N_\text{coll}$ distribution obtained in two different models. Data from Ref.~\cite{Adam:2015jsa}.}
	\label{fig:QpA_dist_Ncoll}
\end{figure}

\begin{figure}[tb]
	\centering
	\includegraphics[scale=\figscale]{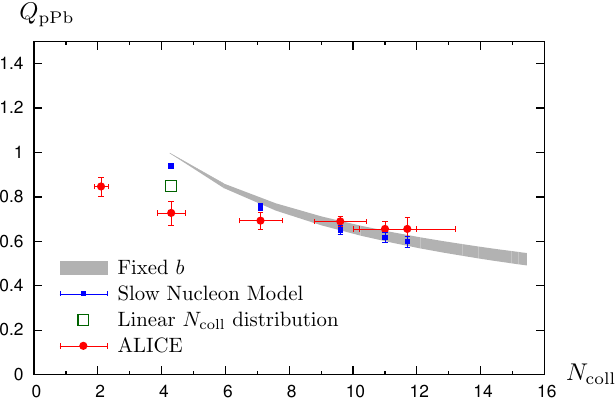}
	\caption{Nuclear modification factor $Q_\text{pPb}$ as a function of $N_\text{coll}$ at $\sqrt{s_{NN}}=5$ TeV compared with ALICE data~\cite{Adam:2015jsa}, both when using a fixed impact parameter and when integrating explicitly over $b$.}
	\label{fig:QpA_Ncoll}
\end{figure}

\begin{figure*}[tbp]
	\centering
	\hspace{-0.3cm}\includegraphics[scale=\figscale]{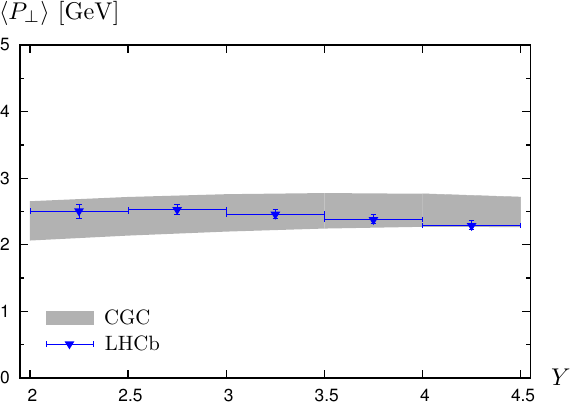}
	\hspace{1.4cm}
	\includegraphics[scale=\figscale]{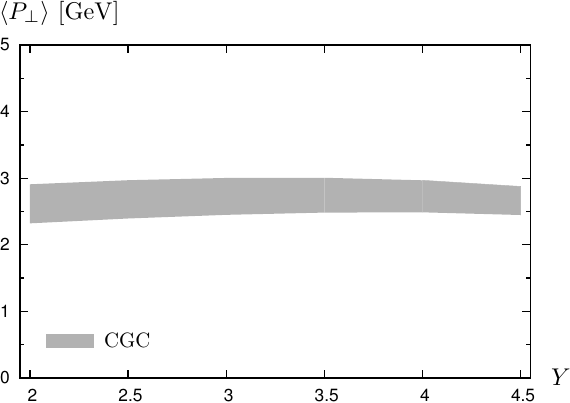}
	\caption{Mean transverse momentum as a function of $Y$ in proton-proton collisions at $\sqrt{s}=7$ TeV (left) and in proton-lead collisions at $\sqrt{s_{NN}}=5$ TeV (right). Data from Ref.~\cite{Aaij:2011jh}.}
	\label{fig:meanpT}
\end{figure*}

\begin{figure*}[tbp]
	\centering
	\includegraphics[scale=\figscale]{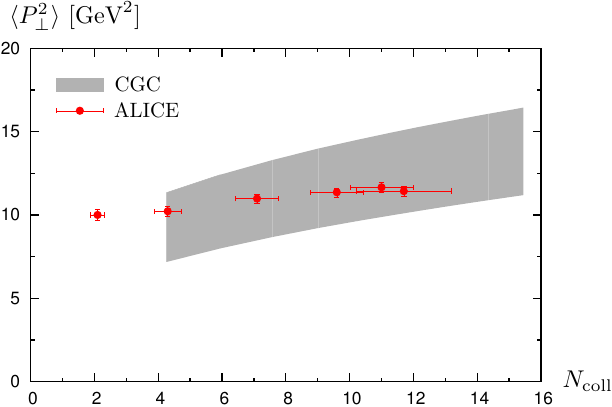}
	\hspace{1cm}
	\includegraphics[scale=\figscale]{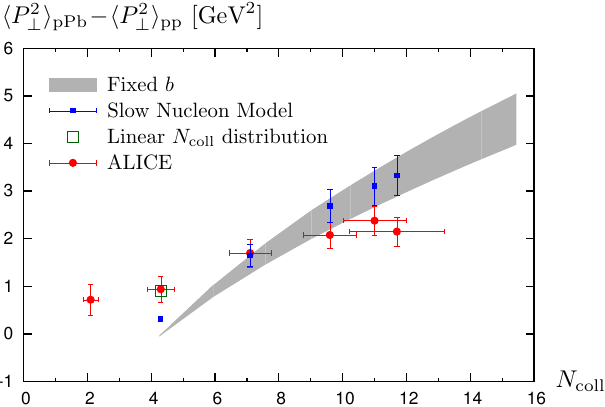}
	\caption{Average transverse momentum squared 
	  $\langle P_\perp^2 \rangle_\text{pPb}$ (left) and nuclear transverse momentum broadening $\langle P_\perp^2 \rangle_\text{pPb}-\langle P_\perp^2 \rangle_\text{pp}$ (right) as a function of $N_\text{coll}$ at $\sqrt{s_{NN}}=5$ TeV compared with ALICE data~\cite{Adam:2015jsa}. For the latter both the result using a fixed impact parameter and when integrating explicitly over $b$ are shown.}
	\label{fig:pT2_Ncoll}
\end{figure*}

\subsection{Explicit integration over the impact parameter}
\label{sec:b_dist}

The results we have shown in the previous section were obtained with a fixed impact parameter chosen so that the number of binary collisions in the optical Glauber model is equal to the average number of binary collisions estimated by ALICE in each centrality class. However, the nuclear modification ratio in a given centrality bin receives contributions from a distribution of different impact parameters. One could therefore argue that having a profile in the impact parameter space and integrating over $b$ could lead to different results. To quantify this effect we will here use two different kinds of $N_\text{coll}$ distributions to obtain distributions in the impact parameter space. The first one is provided by ALICE~\cite{alicecent} and is obtained from the Slow Nucleon Model (SNM)~\cite{Adam:2014qja}. Since in this model the average number of binary collisions is not the same as the one obtained in the hybrid method used in Ref.~\cite{Adam:2015jsa}, we shift the distributions so that $\langle N_\text{coll} \rangle$ matches the one in the third column of Table~\ref{tab:Ncoll}. It should be noted that, contrary to the hybrid method, this method is biased~\cite{Adam:2014qja}. In addition, this is only one possible way of extracting $N_\text{coll}$ distributions at experiments. Other methods could yield significantly different distributions. To try to quantify the dependence of our results on the particular shape of the $N_\text{coll}$ distributions, we will also use, for the 60-80\% centrality bin which is the most sensitive to fluctuations, a simple linearly decreasing distribution. The two parameters of this distribution, its height at the origin ($h$) and the $N_\text{coll}$ value at which it vanishes ($N_\text{max}$), are determined by imposing that it is normalized to unity and that $\langle N_\text{coll} \rangle=\langle N_\text{coll} \rangle_\text{ALICE}$: $N_\text{max}=3 \langle N_\text{coll} \rangle$, $h=2/N_\text{max}$.

In Fig.~\ref{fig:QpA_dist_Ncoll} we show the values obtained for the nuclear modification factor as a function of $P_\perp$ in the 60-80\% centrality bin, both when using a fixed impact parameter and when integrating explicitly over $b$ using the two $N_\text{coll}$ distributions described previously (SNM and linear). The explicit integration over $b$ leads to a smaller $Q_\text{pPb}$ at small transverse momentum. The effect is more pronounced with the linear distribution. Similar results are obtained when looking at $Q_\text{pPb}$ integrated over $P_\perp$ as a function of $N_\text{coll}$, as show in Fig.~\ref{fig:QpA_Ncoll}. In particular, we see that the value of $Q_\text{pPb}$ in the 60-80\% centrality bin obtained with the linear $N_\text{coll}$ distribution is significantly closer to the ALICE data point. On the other hand, the value obtained with the SNM distribution is very close to the fixed impact parameter result.

In conclusion, it is not possible for now to directly compare our impact parameter dependent results with the centrality dependent measurement performed by ALICE. Indeed, for this one would need to have access to an unbiased determination of the $N_\text{coll}$ distributions in each centrality bin, which does not exist at the moment. Here we have tried to estimate the importance of this effect by using $N_\text{coll}$ distributions obtained in two models. The fact that these two models lead to significantly different results for peripheral collisions while the central bins are much less sensitive to fluctuations means that the variation of $Q_\text{pPb}$ as a function of centrality is too model dependent to have a reliable comparison with experimental data.

\begin{figure*}[tb]
	\centering
	\includegraphics[scale=\figscale]{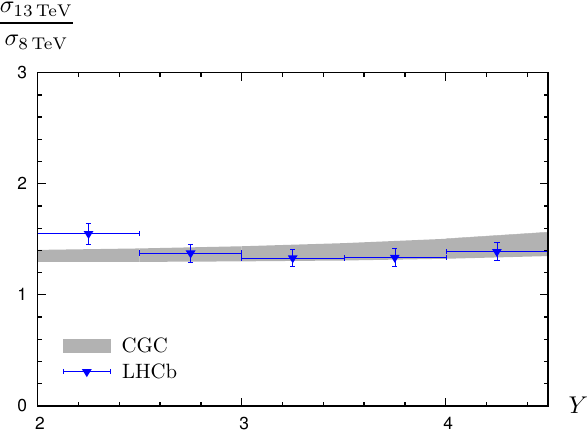}
	\hspace{0.5cm}
	\includegraphics[scale=\figscale]{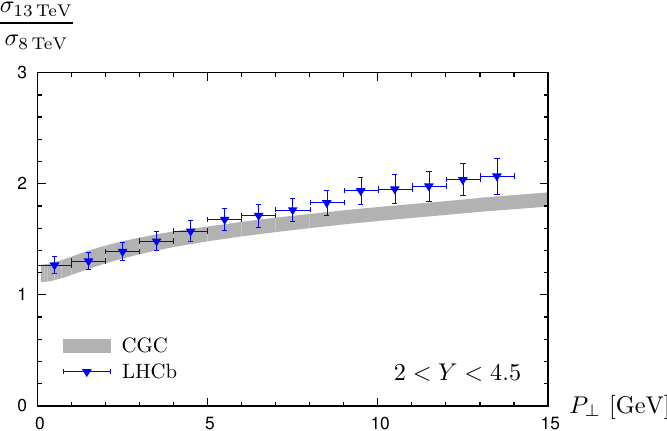}
	\caption{Ratio of the proton-proton cross section at $\sqrt{s}=13$ and 8 TeV as a function of $Y$ (left) and $P_\perp$ (right) compared with LHCb data~\cite{Aaij:2015rla}.}
	\label{fig:ratio_13TeV_8TeV}
\end{figure*}

\begin{figure*}[tb]
	\centering
	\includegraphics[scale=\figscale]{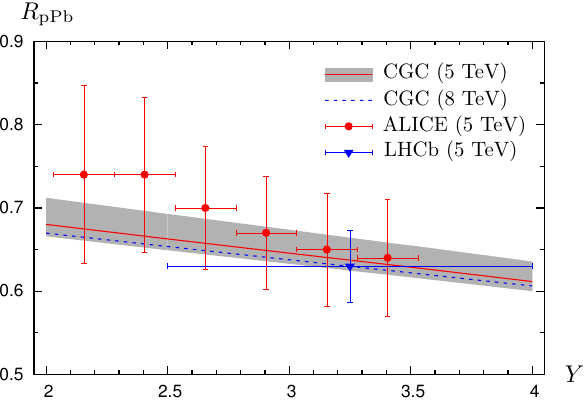}
	\hspace{0.5cm}
	\includegraphics[scale=\figscale]{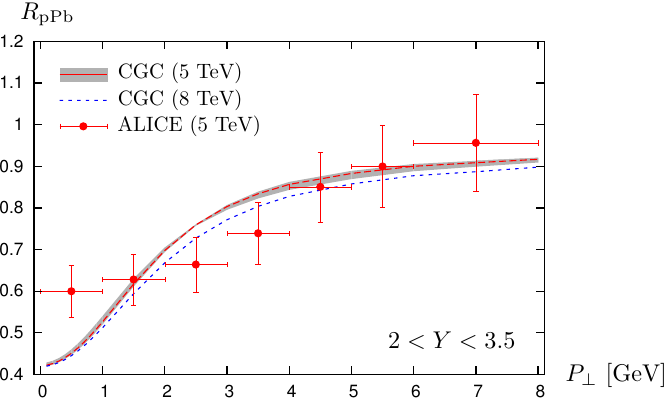}
	\caption{Nuclear modification factor $R_\text{pPb}$ at $\sqrt{s_{NN}}=5$ and 8 TeV as a function of $Y$ (left) and $P_\perp$ (right). Data from Ref.~\cite{Abelev:2013yxa,Aaij:2013zxa,Adam:2015iga}}
	\label{fig:RpA_8TeV}
\end{figure*}

\section{Mean transverse momentum}

In Ref.~\cite{Ducloue:2015gfa} we found that the uncertainty on cross sections both in proton-proton and proton-nucleus collisions was rather large. This uncertainty mostly affects the normalization and therefore quantities such as the nuclear modification factor show a smaller uncertainty. Besides the nuclear modification factor, another observable which is not sensitive to the absolute normalization of the cross section is the mean transverse momentum of the produced $J/\psi$ meson. In Fig.~\ref{fig:meanpT}~(L) we show this quantity as a function of the rapidity in proton-proton collisions at a center of mass energy of 7 TeV and compare with LHCb data~\cite{Aaij:2011jh}. We observe that our calculation is compatible with the data but it is still affected by a relatively large uncertainty. In the collinear approximation on the proton side that we are using here, the mean transverse momentum increases slightly with rapidity, a trend not seen in the data. One must, however, keep in mind that towards central rapidities the intrinsic transverse momentum from also the proton should increase, leading to the opposite behavior. A matching between the collinear and $k_T$-factorized approximations required to fully quantify this effect is beyond the scope of this paper. In Fig.~\ref{fig:meanpT}~(R) we show the same quantity in proton-lead collisions at a center of mass energy $\sqrt{s_{NN}}=5$ TeV.

The ALICE collaboration has also presented results for $\langle P_\perp^2 \rangle$ as a function of $N_\text{coll}$. On Fig.~\ref{fig:pT2_Ncoll}~(L) we see that our calculation agrees with this measurement within the rather large uncertainty band (except for most peripheral collisions, where our calculation is not applicable as explained previously).
When one considers the difference between $\langle P_\perp^2 \rangle$ in proton-lead and in proton-proton collisions, as shown on Fig.~\ref{fig:pT2_Ncoll}~(R), the uncertainty on our calculation shrinks and shows a too strong variation as a function of $N_\text{coll}$, both when using a fixed impact parameter and when integrating explicitly over $b$ using the $N_\text{coll}$ distributions obtained in the Slow Nucleon Model. However, as in section~\ref{sec:b_dist}, the results obtained when integrating over $b$ depend strongly on the exact shape of the $N_\text{coll}$ distributions used. In particular, one can see that using a linear $N_\text{coll}$ distribution leads to a better agreement with data for peripheral collisions.

\section{Dependence on the center of mass energy}

\subsection{Proton-proton collisions}

In this work we use the simple color evaporation model to describe the hadronization of $c\bar{c}$ pairs into $J/\psi$ mesons. The normalization of cross sections then depends on a non perturbative constant $F_{J/\psi}$, see (\ref{eq:dsigmajpsi}). The uncertainty associated with this parameter can be eliminated by studying the ratio of cross sections at different center of mass energies. In addition, from the experimental point of view, systematic uncertainties can cancel to some extent in this ratio. Such a measurement has been made possible at the LHC for proton-proton collisions thanks to the recent increase of $\sqrt{s}$ from 8 to 13 TeV. In particular the ratio $\sigma_{13 \, \text{TeV}}/\sigma_{8 \, \text{TeV}}$ was studied as a function of $Y$ and $P_\perp$ by the LHCb collaboration~\cite{Aaij:2015rla}.
In Fig.~\ref{fig:ratio_13TeV_8TeV} we compare these data with the results that we obtain for this ratio in our model. The resulting uncertainty is rather small and the agreement with data is quite good, in particular at large rapidity and relatively low transverse momentum which is the kinematical domain where our calculation is expected to be the most reliable.

\subsection{Proton-nucleus collisions}

Thanks to its recent upgrade, the LHC may also perform proton-lead collisions at a higher center of mass energy in the future. Here we study how our results would be affected by a change of $\sqrt{s_{NN}}$ from 5 to 8 TeV. In Fig.~\ref{fig:RpA_8TeV} we show the nuclear modification factor $R_\text{pPb}$ at forward rapidity as a function of $Y$ and $P_\perp$ at these two energies, as well as existing LHC data at $\sqrt{s_{NN}}=5$ TeV. The values at $\sqrt{s_{NN}}=5$ TeV shown here differ slightly from the ones shown in Figs.~8 and 10 of Ref.~\cite{Ducloue:2015gfa} because we corrected a numerical problem which was causing the region of large impact parameters (where the saturation scale of the lead nucleus falls below the one of the proton, see Fig.~\ref{fig:Qs2_Ncoll}~(L)) to be neglected. Here we impose $R_\text{pA}=1$ in this region, as in Ref.~\cite{Lappi:2013zma}. As one could expect, the higher center of mass energy leads to a stronger suppression due to the higher densities reached in the target. However, the effect is quite small, in particular compared to the size of the uncertainties. For this reason we only show, for $\sqrt{s_{NN}}=8$ TeV, our results for the ``central'' values of the parameters ($m_c=1.29$ GeV and $Q=\sqrt{\P_\perp^2+M^2}$). For  $\sqrt{s_{NN}}=5$ TeV we show both the central value and the uncertainty band corresponding to the variation of $m_c$ and $Q$. Therefore, while measuring forward $J/\psi$ production in 8 TeV proton-lead collisions could help reduce experimental uncertainties by getting rid of the interpolation needed for the proton-proton reference, we do not expect significantly stronger nuclear effects at this energy. This is not surprising since we use the same dipole cross sections as in Ref.~\cite{Lappi:2013zma}, where a weak energy dependence of the nuclear modification factor was found in single inclusive particle production.

\section{Conclusions}

In this paper we have extended our study of forward $J/\psi$ production in proton-nucleus collisions in the Color Glass Condensate framework to new kinematics and observables. In particular we have studied the nuclear suppression at negative rapidities by describing the nucleus probed at large $x$ in terms of nuclear parton distribution functions. We achieved a quite good description of experimental measurements of this quantity, even if the uncertainty is larger at backward than at forward rapidities because the nuclear PDFs are not yet very strongly constrained by data. This allowed us to compute the forward to backward ratio, again with a good agreement with data within the rather large uncertainties. We have also studied the centrality dependence of our calculation. While using the optical Glauber model to extend the description from a proton target to a nucleus leads to a better agreement with experimental data for minimum bias observables than previous calculations in the same framework, it is difficult to compare directly the resulting centrality dependence to experimental data. Indeed, using a fixed impact parameter obtained in the optical Glauber model from the average number of binary collisions estimated by ALICE leads to a too strong centrality dependence. On the other hand, to integrate explicitly over the impact parameter one has to use $N_\text{coll}$ distributions based on various assumptions and our calculation is very sensitive to the exact shape of these distributions. Finally we have studied how the nuclear modification factor would be affected by an increase of the center of mass energy achievable at the LHC. As expected nuclear effects are stronger but the change is too small to be significant given the size of theoretical uncertainties.

\section*{Acknowledgments}
T.~L. and B.~D. are supported by the Academy of Finland, projects
267321 and 273464. 
H.~M. is supported under DOE Contract No. DE-SC0012704.
This research used computing resources of 
CSC -- IT Center for Science in Espoo, Finland.
We would like to thank C. Hadjidakis and I. Lakomov for 
discussions on the ALICE data.

\providecommand{\href}[2]{#2}\begingroup\raggedright\endgroup

\end{document}